\documentclass[12pt,twoside]{article}
\usepackage[dvips]{graphicx}
\usepackage{amssymb, amsmath}           
\usepackage{citesort}                   
\usepackage{ulem,fancyheadings}
\usepackage[indent]{caption2}

\makeatletter
\renewcommand\@openbib@code{
     \advance\leftmargin\bibindent
     \itemindent -\bibindent
     \listparindent \itemindent
     \parsep -0.8ex 
     }
\renewcommand\section{\@startsection {section}{1}{\z@}%
                                   {-4.3ex \@plus -.2ex \@minus -.2ex}%
                                   {2.1ex \@plus.2ex}%
                                   {\normalfont\underline}}
\renewcommand\subsection{\@startsection{subsection}{2}{\z@}%
                                     {-2.1ex\@plus .1ex \@minus -.1ex}%
                                     {2.1ex \@plus .2ex \@minus -.1ex}%
                                     {\normalfont\underline}}
 \makeatother
 
\begin{document}
 \pagestyle{fancy}
 \setlength{\headrulewidth}{0pt}
 \lhead{}
 \rhead{\begin{picture}(0,0)\put(120,70){\it\thepage}\end{picture}}
 \chead{}
 \cfoot{} 
 \hphantom{.}\vskip10mm
\begin{center}
{Linear and non-linear dielectric spectroscopy\\ on ammonium doped Rochelle salt }
\end{center}
\vskip10.5mm
\begin{center}
 U. Schneider, P. Lunkenheimer, J. Hemberger and A. Loidl\\
 Experimentalphysik V, Universit\"{a}t Augsburg, D-86135 Augsburg, Germany
\end{center}
\vskip10.5mm

 \noindent
\underline{Abstract} In this work we present a thorough investigation of the mixed system
of Rochelle salt doped with ammonium, with $0\leq x\leq 1$ for
[Na$($K$_{1-x}($NH$_4)_x)$C$_4$H$_4$O$_6\cdot 4$H$_2$O]. This interesting system is known
to exhibit a rich phase diagram with a variety of polar and non-polar phases. However,
most of the earlier investigations date back many decades and the experimental advances
made since then should enable new insight into the physical nature of its phase diagram.
We studied single crystals of 18 doping concentrations using linear and {\it
non}\,-linear frequency-dependent dielectric response spectroscopy. In the low field
measurements a variety of relaxation features are detected and analyzed in detail. The
results for two concentrations of $x=0$ and $x=0.15$ are described qualitatively with a
simple model. Possibly the most important outcome of the present work is the detection of
a new electrically ordered phase for concentrations $0.18<x<0.89$.
 \vskip 10mm

 \vskip 14mm

 \vskip 14mm



\section{INTRODUCTION}
Rochelle salt (NaK\,C$_4$H$_4$O$_6\cdot{}4$H$_2$O) was first synthesized in 1655 by the
pharmacist {\it Pierre Seignette} in La Rochelle (France). Therefore the name {\it
Seignette salt} is also in use in the literature. Although it enjoyed a rapid
distribution for medical and chemical purposes \cite{Mac77} from the very beginning, its
unusual physical properties, especially the large piezoelectric coefficient, were
discovered in the late 18$^{\rm th}$-century only \cite{Poc06}. Its extraordinary
ferroelectric properties along the crystallographic {\it a}\,-axis were revealed even
later by Valasek \cite{Val21} who transferred the terminology of magnetic phase
transitions ({\it Curie}\,-temperature) into the physics of dielectric materials. In his
work he found two second-order phase transition temperatures at 297\,K and 255\,K which
indicate a narrow temperature range where ferroelectric (FE) order exists.

As reported in literature \cite{kourtschatov36,takagi58,makita58,jona62,sandy67}, on
doping the pure compound, NaK-tartrate\footnote{The name {\it tartrate} is derived from
french {\it cr\`eme de tartre} for the crystalline furring deposit in containers of
wine.}, with ammonium, the temperature range of the FE phase is reduced till the order is
fully suppressed at approximately $x=0.025$. Mixed crystals up to $x \approx 0.18$ remain
paraelectric (PE) down to lowest temperatures. For ammonium concentrations $0.18<x<0.89$
there is a FE phase below a second order transition which was reported to prevail down to
lowest temperatures \cite{jona62}. Between $x=0.89$ and the pure NaNH$_4$-tartrate
compound this FE phase is no longer present and a new phase appears with a first order
phase transition. This phase shows a polar order that cannot be reversed by electrical
fields $E \leq 20$\,kV/cm but by applying additional mechanical stress.

The crystal structure of the non-polar phases of pure NaK-tartrate was determined to be
orthorhombic ($P2_12_12$) \cite{beevers41}. The structure of the polar phase was found to
be monoclinic ($P2_111$) \cite{frazer54}. Despite numerous theoretical approaches there
is no overall consensus concerning the explanation of the strange re-entrant behaviour.
To enhance our understanding of this puzzling feature we were especially interested in
investigating the dipolar reorientation in the low temperature PE phase below the second
phase transition of the pure NaK-tartrate and the doping range of ammonium with
$0.025<x<0.18$ between the two FE phases. We applied both ordinary broadband dielectric
spectroscopy to examine the frequency-dependent linear response and high-field
spectroscopy to detect the non-linear hysteretic response. Various samples of 18
different doping concentrations, were investigated covering all characteristic regions of
the phase diagram.

Finally we will show that the simple model of Blinc {\it et al.} \cite{blinc72} can be
used to describe the doping range of $0<x<0.18$ qualitatively.

 \chead[U. Schneider, P. Lunkenheimer, J. Hemberger and A. Loidl]
 {Linear and non-linear dielectric spectroscopy...}

\section{EXPERIMENTAL DETAILS}

The crystals were grown from an aqueous solution. The crystallographic axes were
identified according to the geometrical properties reported in literature ({\it e.g.}
\cite{jona62,mason50}). The average size of the raw crystals was about 1\,cm$^3$. They
were polished down to a thickness along the {\it a}\,-axis $\approx{}1$\,mm with an area
in the {\it b-c}\,-plane of $\approx 100-150$\,mm$^2$. For applying dielectric
spectroscopy to these samples, both faces were covered with silver paint.

We used commercially available closed-cycle $^4$He refrigerators for cooling. The
frequency-dependent linear dielectric measurements were performed with an autobalance
bridge (HP4284) using pseudo 4-point technique. The frequency range covered is
$20$\,Hz\,$\leq\nu\leq$\,$1$\,MHz. The high-field non-linear measurements were made with
a home-made setup \cite{hem94} designed according to Sawyer and Tower \cite{sawyer30}.

\section{RESULTS AND DISCUSSION}

\begin{figure}[htbp]
\begin{center}
\includegraphics[clip,width=9.5cm]{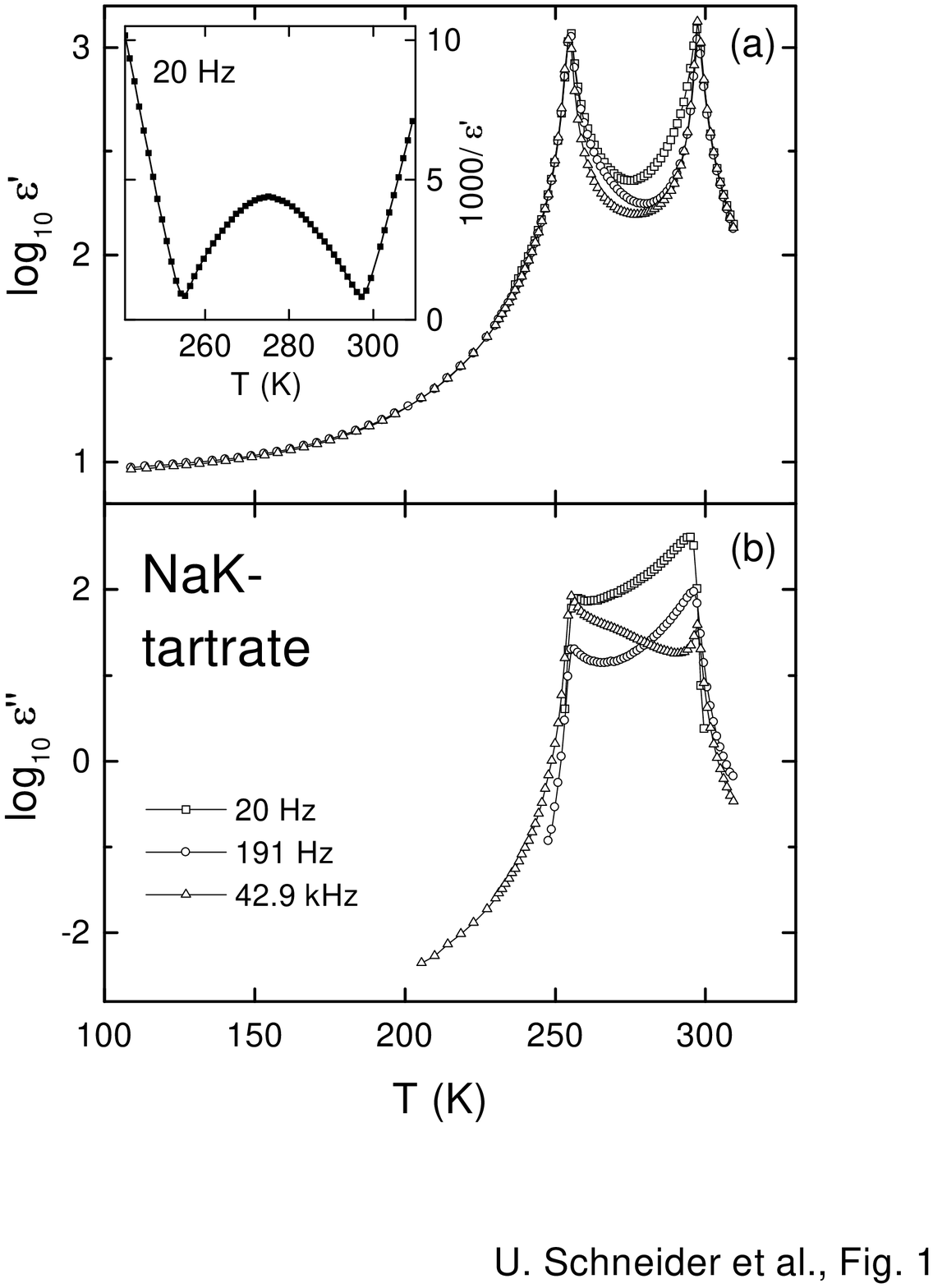}
\caption{Temperature dependence of real (a) and imaginary part (b) of the dielectric
permittivity of NaK-tartrate for three frequencies. The inset shows a {\it Curie}\,-plot
of $\varepsilon$'$(T)$ for $\nu = 20$\,Hz. The lines are drawn to guide the eye.}
\label{fig:lin000}
\end{center}
\end{figure}

Frequency-dependent linear measurements were performed in the whole ammonium doping range
of $0\leq x\leq 1$ for the filed along the crystalline $a$ axis. The various PE and FE
phases showing up in the phase diagram are discussed in detail below. In the following,
the presentation of the spectral results will be restricted to certain concentrations,
which exhibit dielectric properties, typical for the respective phases. In
Figure~\ref{fig:lin000}
we present the temperature dependent data of both, the real and the imaginary part of the
dielectric function, $\varepsilon^*=\varepsilon$'$-i\varepsilon$'', of pure NaK-tartrate
for three frequencie s. Clearly the two second-order phase transitions at
$T_{c1}=296.6$\,K and $T_{c2}=254.6$\,K appear as sharp anomalies in $\varepsilon$'$(T)$
(Figure~\ref{fig:lin000}a). In the temperature range of the ferroelectric phase for both,
$\varepsilon$' and $\varepsilon$'' (Figure~\ref{fig:lin000}b) there are indications of
relaxational behaviour which, however, cannot be analyzed further due to the underlying
strong temperature dependence in the vicinity of the phase transitions. Moreover, for
$\varepsilon$'' one can observe unusually high losses due to domain wall movement. The
inset in Figure~\ref{fig:lin000}a shows the inverse of $\varepsilon$' for $\nu = 20$\,Hz.
This plot reveals Curie-Wei{\ss} behaviour at $T<T_{c1}$ and $T>T_{c2}$ while at
$T_{c1}<T<T_{c2}$ its detection is hampered by the relatively close vicinity of both
phase-transition temperatures and the relaxation behaviour mentioned above. The
re-entrant phase transitions, PE$\rightarrow$FE$\rightarrow$PE, prevail up to an ammonium
concentration of $0.025$.

\begin{figure}[htbp]
\begin{center}
\includegraphics[clip,width=9.5cm]{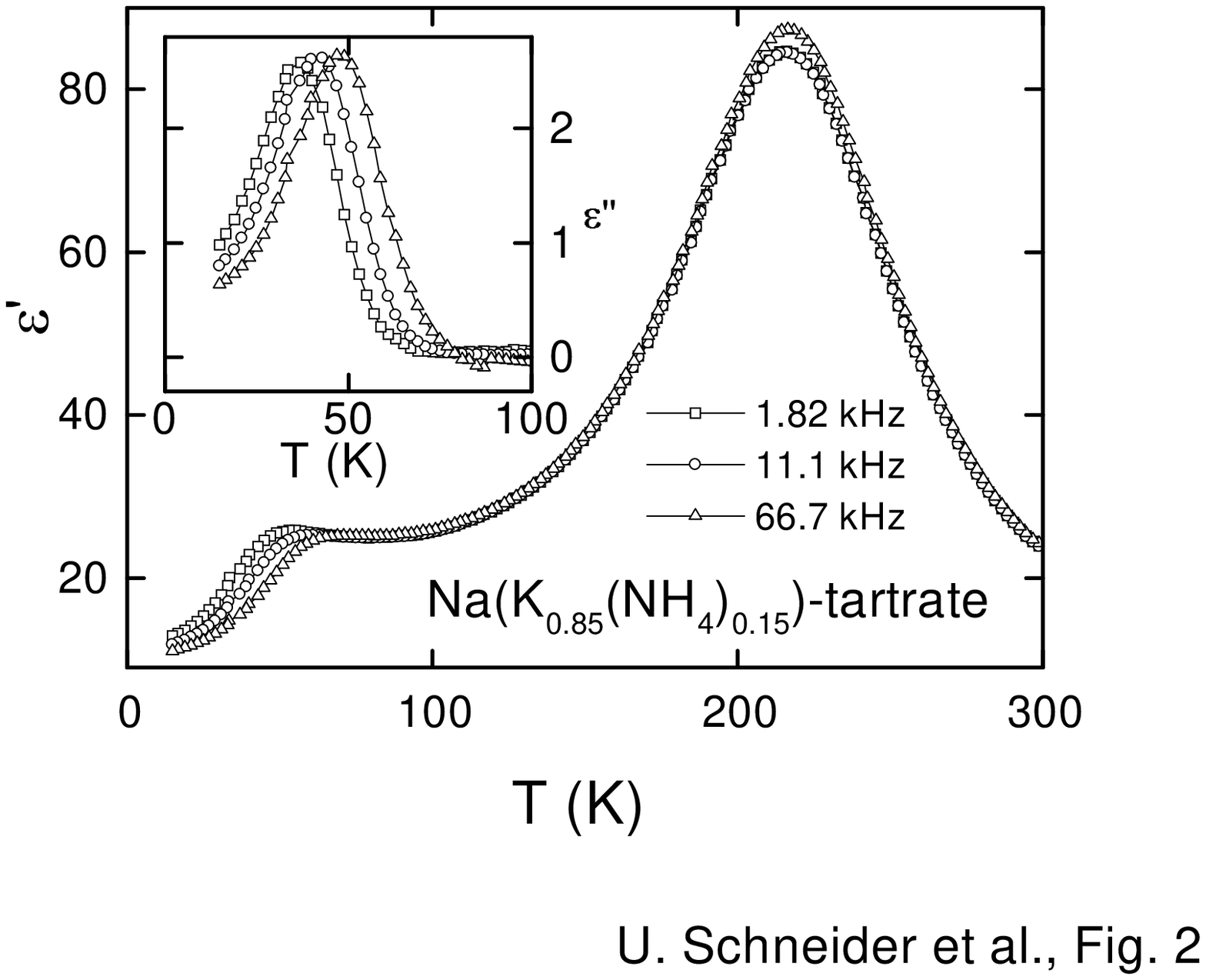}
\caption{Temperature dependence of the real part of the dielectric function of
Na$($K$_{0.85}($NH$_4)_{0.15})$-tartrate for three frequencies. The inset shows the
imaginary part below 100\,K. The lines are drawn to guide the eye.} \label{fig:lin015}
\end{center}
\end{figure}

As an example for the doping range of $0.025<x<0.18$ we present the data for $x=0.15$ in
Figure~\ref{fig:lin015}.
In this range there is no phase transition. The FE correlations, which produce the
ordered phase for doping levels below $2.5$\,\%, are reduced below a critical limit (see
section~\ref{sec:theory}). In Figure~\ref{fig:lin015} the real part $\varepsilon$'$(T)$
shows a relatively flat, broad peak at about 217\,K, but no FE ordering is detected in
the nonlinear results (not shown), which reveal simple PE loss-loops. At low
temperatures, between 10 and 90\,K there is a pronounced relaxation feature. It shows up
as sigmoidally shaped curves in $\varepsilon$'$(T)$ and as peaks in $\varepsilon$''$(T)$
(inset of Figure~\ref{fig:lin015}), both shifting to higher temperatures with increasing
frequency. The temperature evolution of the corresponding relaxation times will be
discussed below.

\begin{figure}[htbp]
\begin{center}
\includegraphics[clip,width=9.5cm]{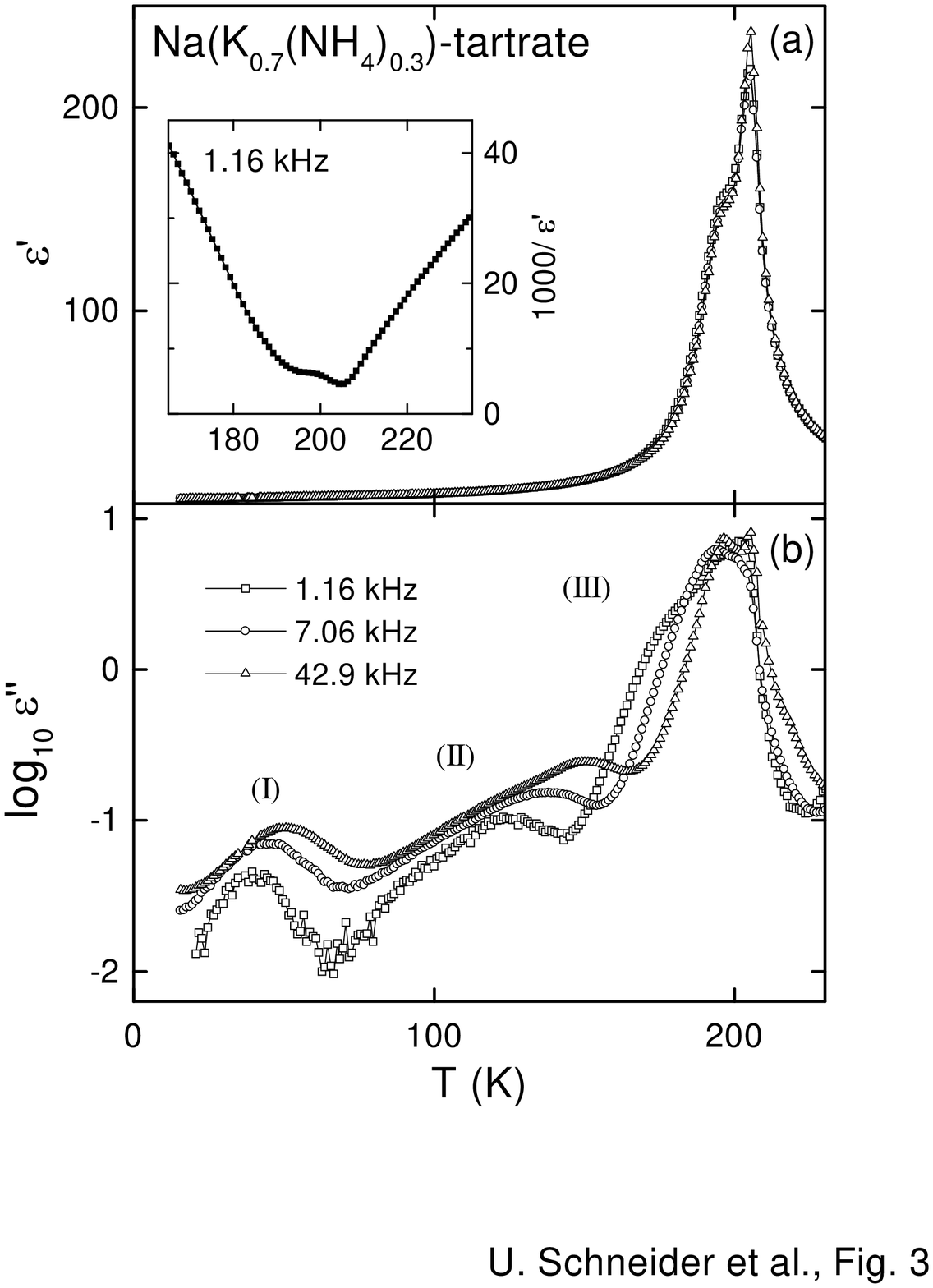}
\caption{Temperature dependence of real (a) and imaginary part (b) of the dielectric
permittivity of Na$($K$_{0.70}($NH$_4)_{0.30})$-tartrate for three frequencies. The inset
shows a {\it Curie}\,-plot of $\varepsilon$'$(T)$ for $\nu = 1.16$\,kHz. The numbers (I),
(II), and (III) denote the three observed dominant relaxation r\'egimes. The lines are
drawn to guide the eye.} \label{fig:lin030}
\end{center}
\end{figure}

The doping range $0.18<x<0.89$ was reported to exhibit a ferroelectrically ordered phase
\cite{kourtschatov36,takagi58,makita58,jona62,sandy67}. All samples investigated in this
range show qualitatively the same characteristics.
Figure~\ref{fig:lin030}
shows the data of a typical crystal (30\,\% doping) within this range. Like the pure
NaK-tartrate this sample shows a sharp peak in $\varepsilon$'$(T)$
(Figure~\ref{fig:lin030}a) due to a second-order transition at 205\,K. On closer
inspection of the data a shoulder on the low temperature side of this peak can be
identified at 187\,K, which indicates a second phase transition. This feature is even
more pronounced in the inverse representation shown in the inset. This additional
transition is detected in all data on crystals in this doping range. The imaginary part
in Figure~\ref{fig:lin030}b shows at least three relaxation processes, namely in the
temperature ranges of $10-90$\,K (I), $90-180$\,K (II) and $180-200$\,K (III)
superimposed to the transition anomalies. The three relaxations can be clearly marked by
their dispersive behaviour which will be discussed in detail below.

\begin{figure}[htbp]
\begin{center}
\includegraphics[clip,width=9.5cm]{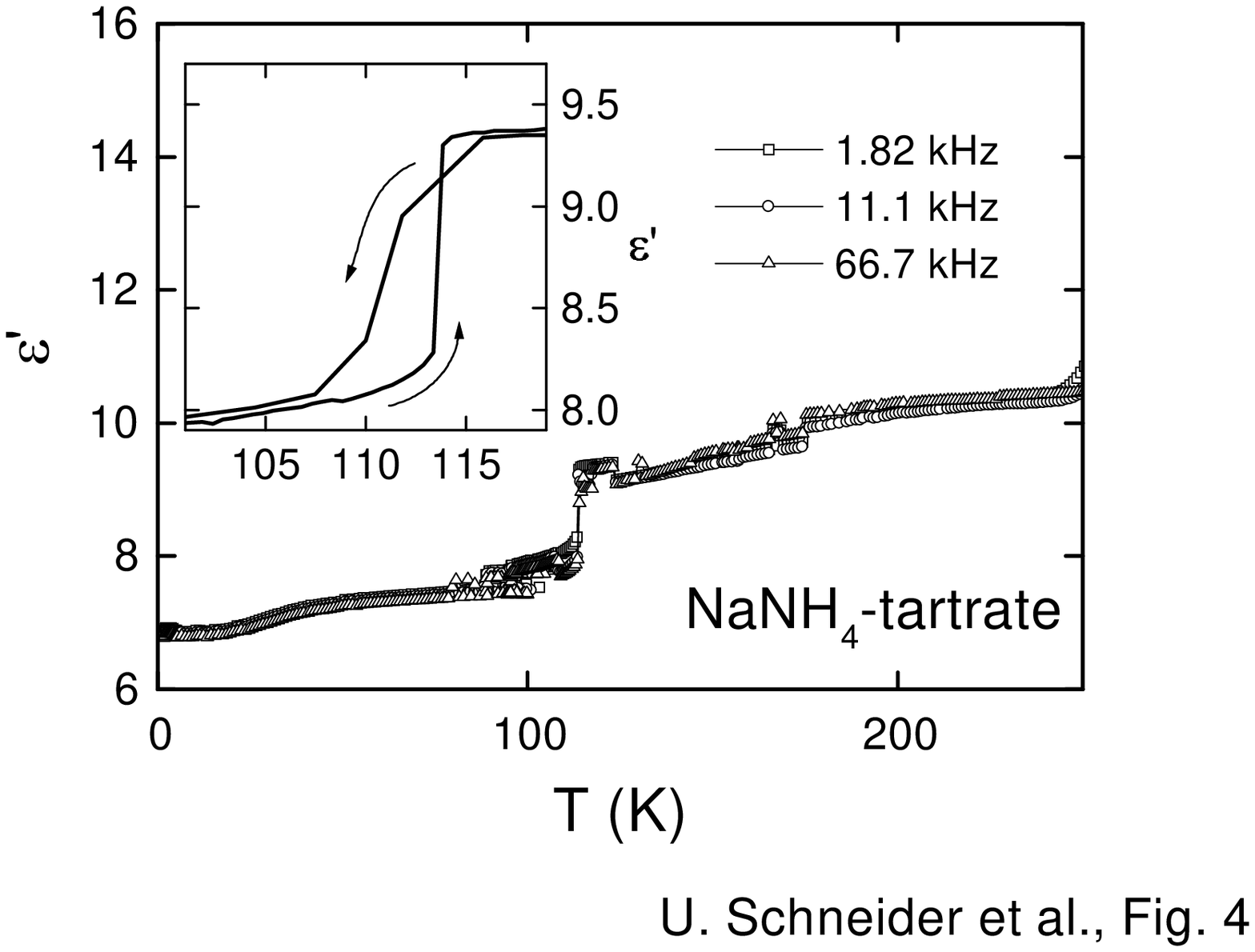} \caption{Temperature dependence of
the real part of the dielectric function of NaNH$_4$-tartrate for three frequencies. The
lines are drawn to guide the eye. The inset shows a magnified view of the first order
phase-transition region.} \label{fig:lin100}
\end{center}
\end{figure}

For $x>0.89$ a new phase transition appears. The data on pure NaNH$_4$-tartrate are
presented in Figure~\ref{fig:lin100}.
The transition at 112\,K from the PE to an ordered phase is of first order as can be
deduced from the hysteretic step in $\varepsilon$'$(T)$ (inset of
Figure~\ref{fig:lin100}).

\begin{figure}[htbp]
\begin{center}
\includegraphics[clip,width=9.5cm]{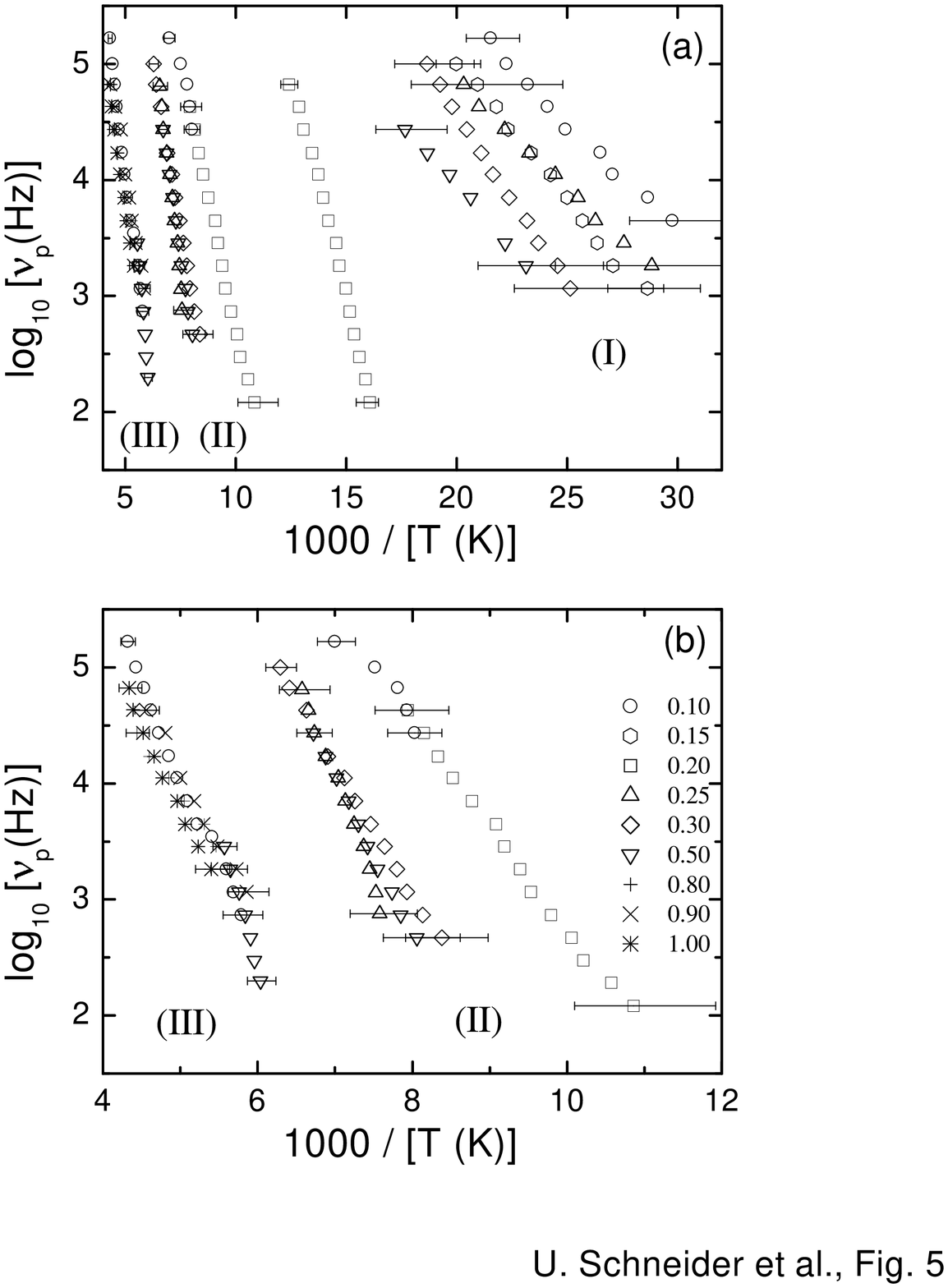}
\caption{Collection of relaxation rates in Arrhenius representation of the processes,
which are observed in the whole doping range; part (b) shows an enlarged view of the high
temperature data of (a).} \label{fig:relax1}
\end{center}
\end{figure}

\begin{figure}[htbp]
\begin{center}
\includegraphics[clip,width=9.5cm]{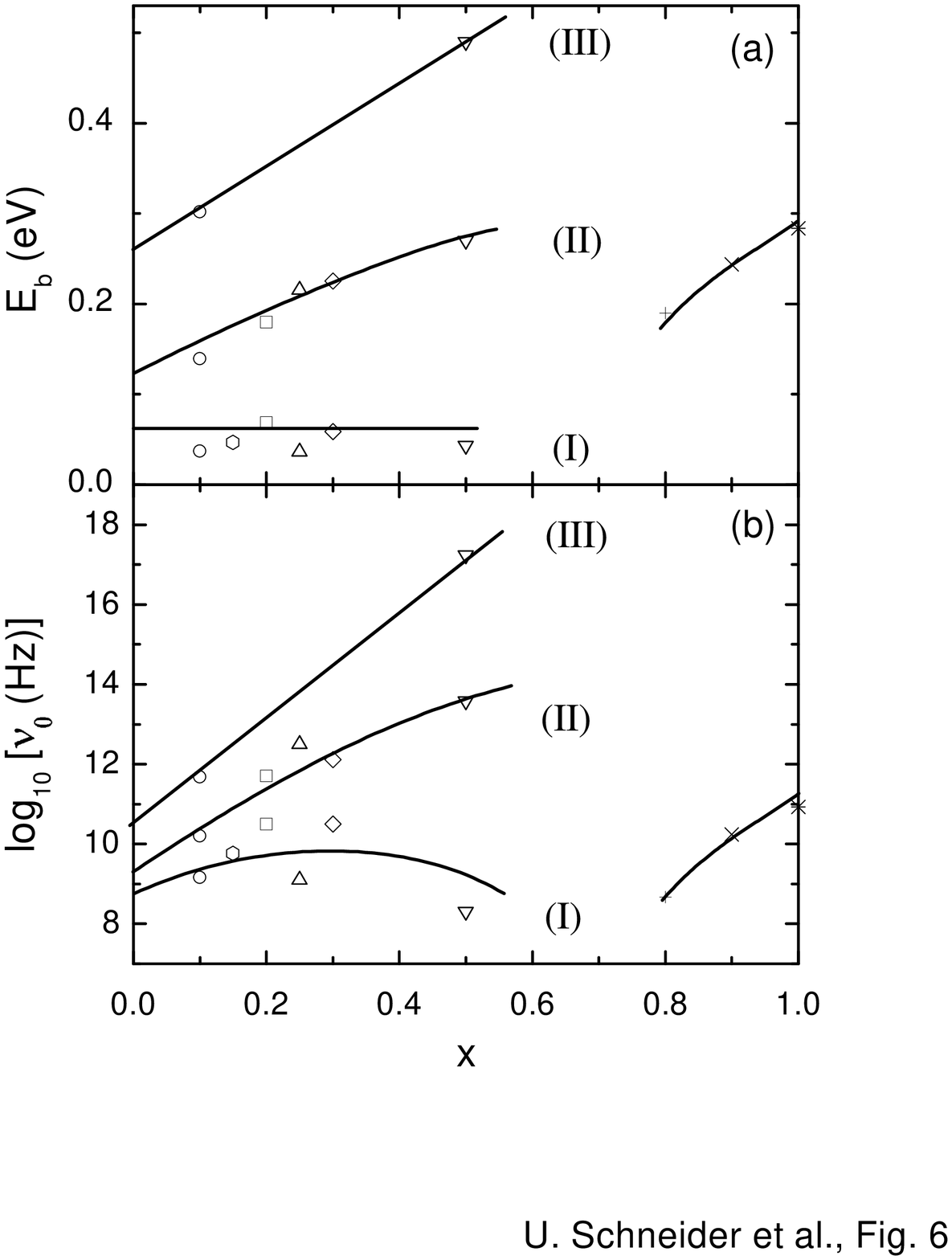}
\caption{Parameters $E_b$ (a) and $\nu_0$ (b) of the {\it Arrhenius}\,-fits of $\nu_p(T)$
(cf. Equation~\protect\ref{eq:Arrhenius}) versus doping concentration $x$. The lines
suggest interpolated doping dependences of the plotted parameters for the three ranges
(I), (II) and (III) classified according to the corresponding temperature ranges.}
\label{fig:relax2}
\end{center}
\end{figure}

We will now focus on the variety of relaxation processes mentioned above. At the
temperature of the peak maximum in $\varepsilon$''$(T)$ the relaxation rate $\nu_p$ can
be estimated to be equal to the measuring frequency. The temperature dependent $\nu_p$,
determined in this way for the collection of evaluated processes is shown in
Figure~\ref{fig:relax1}
in an {\it Arrhenius} representation. One may classify three groups of separate processes
in the inverse temperature ranges of $17-30$ (I), $6-12$ (II) and $4-6$ (III) in units of
1000/K (with the exception of one relaxation in the $x=0.2$ doped samples). Within the
error bars the various data sets can be fitted\footnote{The fits are not shown to improve
readability of the plots.} by straight lines (nearly parallel within one group),
indicating thermally activated behaviour according to
 \begin{equation}
\nu_{p}(T)=\nu_0 \exp\left(-\frac{E_b}{k_BT}\right)\;  .\label{eq:Arrhenius}
\end{equation}
Here $\nu_0$ is the attempt frequency and $E_b$ denotes the energy barrier. The fits
result in similar parameters for each group. In
Figure~\ref{fig:relax2}
we plot the values of these parameters versus the doping concentration $x$. The lines are
drawn to demonstrate the continuous development of the parameters within the three groups
(I), (II) and (III) up to $x\approx 0.6$. The lines, connecting the Arrhenius parameters
of the three concentrations above $x=0.8$, also belong to the group (III) in terms of the
absolute values of the relaxation times, while their Arrhenius parameters deviate
considerably from those of the other members of this group. The involved errors for the
values shown in Figure~\ref{fig:relax2} are about 25\,\% resulting from the error bars
given in Figure~\ref{fig:relax1}. Within these limits all parameters follow the behaviour
sketched by the lines in Figure~\ref{fig:relax2} reasonably well. Presumably the observed
relaxations result from intra-molecular motions, e.g. of the carboxyle groups or the
hydrogen bonds. This finding is (for group I) also corroborated by the relatively small
doping dependence of the plotted parameters. From the extremely high values of the
attempt frequencies of group (III), deviations from Arrhenius behaviour can be expected
at high temperatures, indicating a collective nature of the probed motion. Assuming an
increase of disorder upon doping the width of the associated random bonds distribution is
proportional to a characteristic "freezing" temperature $T_f$ which thus shifts to higher
temperatures for higher doping concentrations. The Arrhenius fit therefore yields a
higher value for $E_b$ in the same temperature interval and an increased overestimation
of $\nu_0$. For all relaxation processes observed, the inspection of
$\varepsilon$''$(\nu)$ (not shown) reveals a half width of the loss peaks much broader
than expected for Debye relaxation. Such behaviour is usually ascribed to a distribution
of relaxation times.

\begin{figure}[htbp]
\begin{center}
\includegraphics[clip,width=9.5cm]{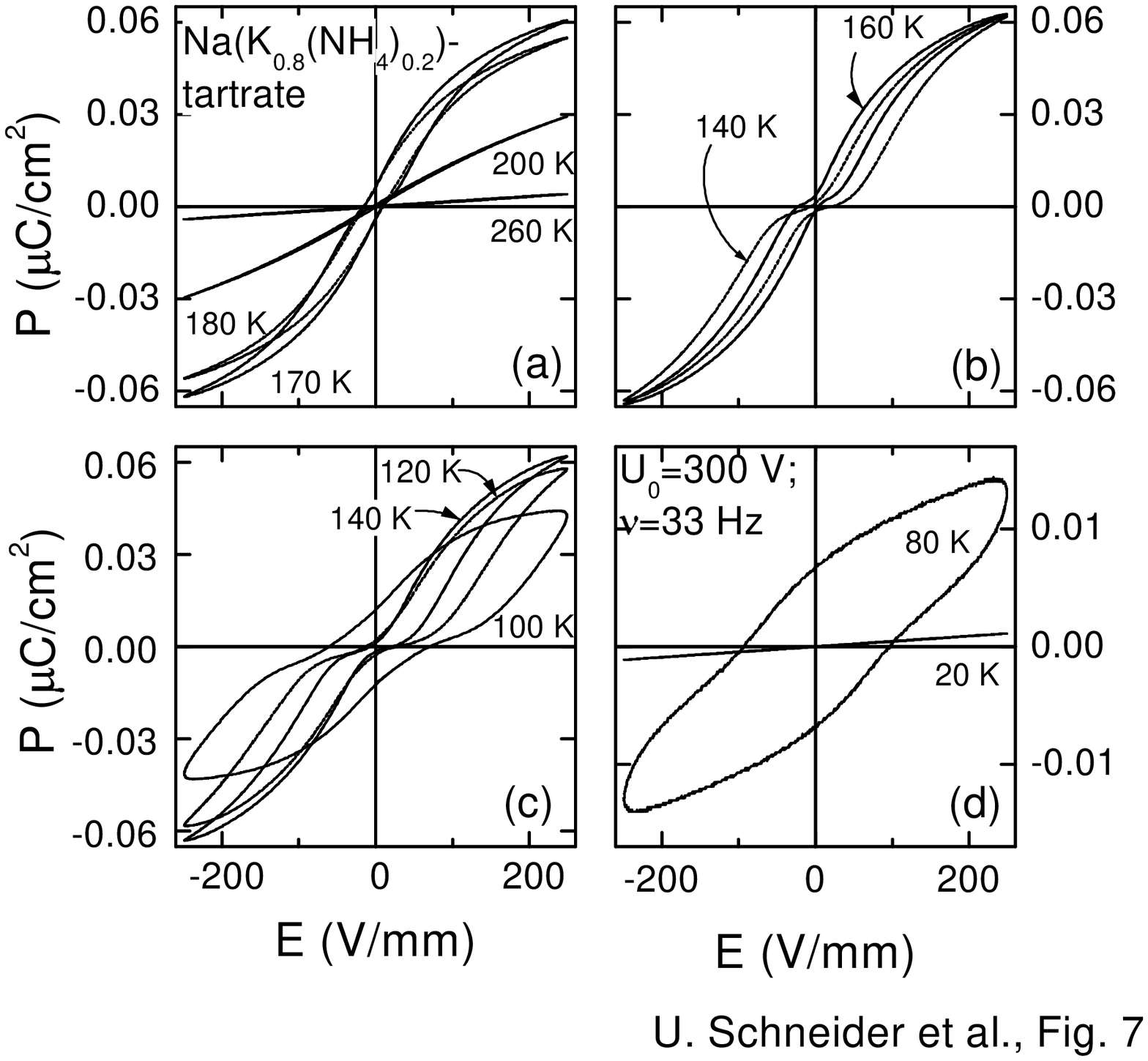}
\caption{Nonlinear response of Na$($K$_{0.80}($NH$_4)_{0.20})$-tartrate for various
temperatures.} \label{fig:nonlinear020}
\end{center}
\end{figure}

In the following we will discuss the high field results of two samples with the doping
concentrations $x=0.2$ and $x=1$. In
Figure~\ref{fig:nonlinear020}
the field dependent polarization for the 20\,\% doped sample is presented for various
temperatures. At temperatures $T>200$\,K linear behaviour is observed. At 200\,K a
hysteresis loop starts to open up, typical for a transition to FE order. Below 160\,K a
frequency independent wasp waisted contraction of the curves near the origin is observed.
This effect can be attributed to an overlay of the present FE order with {\it
anti}\,ferroelectric (AFE) correlations \cite{grindlay70,Haertling87,Xu91}. Another
possible reason for this phenomenon found also in magnetically ordering materials might
be a reorientation of domains about 90$^\circ$ as proposed by Weber {\it et al.} for
cobalt films on copper \cite{Weber96}. However, the shoulder, observed in
$\varepsilon$'$(T)$ as discussed above (Figure~\ref{fig:lin030}a) speaks in favour of the
first interpretation. It clearly indicates a second-order transition. Below about 140\,K
the polarization becomes significantly reduced, which can be ascribed to growing
coercitive fields with decreasing temperature leading to an incomplete reorientation of
the polarization.

\begin{figure}[htbp]
\begin{center}
\includegraphics[clip,width=9.5cm]{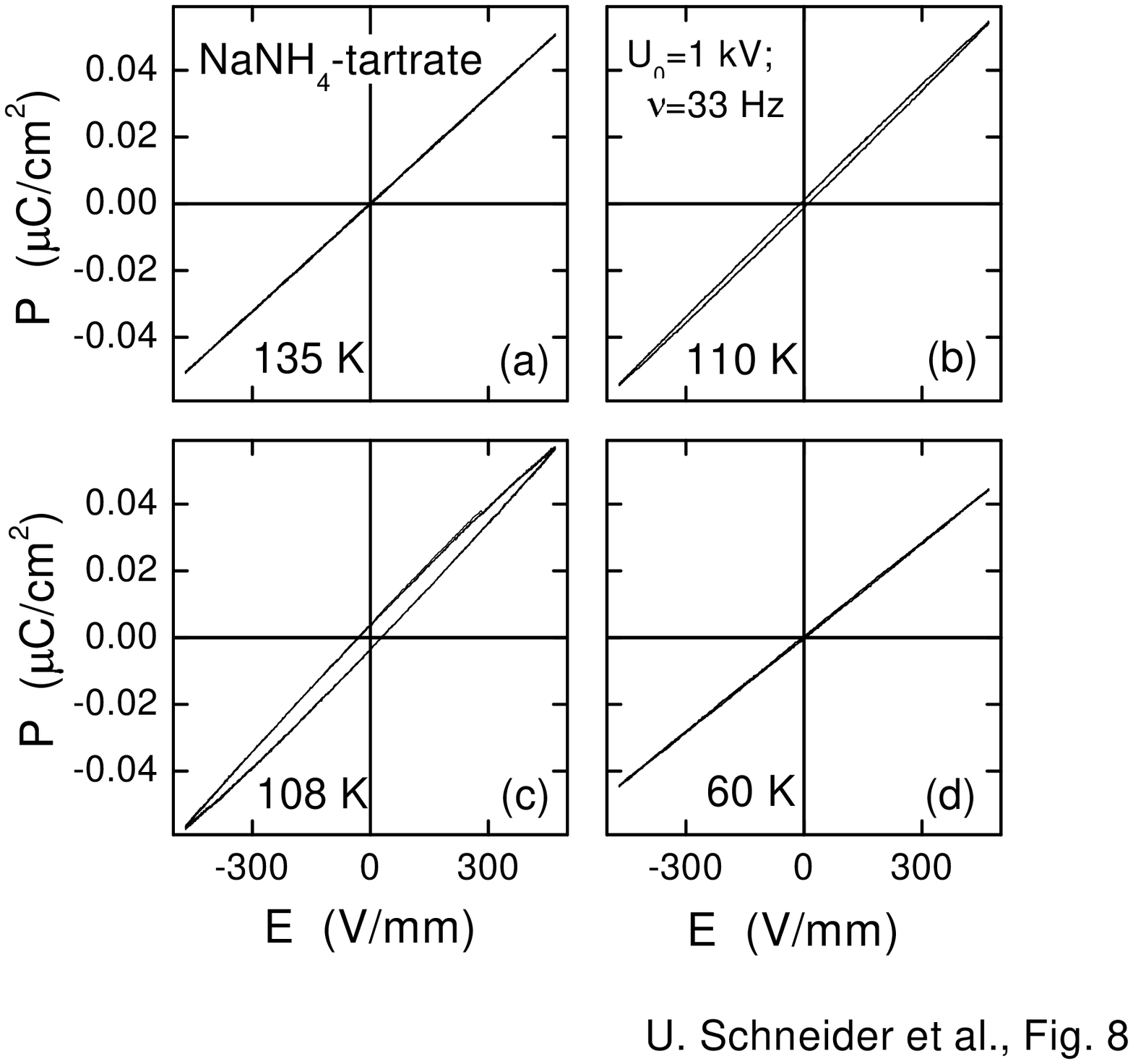}
\caption{Nonlinear response of NaNH$_4$-tartrate for various temperatures.}
\label{fig:nonlinear100}
\end{center}
\end{figure}

Figure~\ref{fig:nonlinear100}
shows the high field response of pure NaNH$_4$-tartrate: even with the highest applied
fields (just below the threshold limit) no electrical ordering can be introduced. An
abrupt change to ellipsoidally shaped $P(E)$ loops can be observed around the temperature
of the first-order phase transition at 112\,K indicating an enhanced dielectric loss in
this region.

\begin{figure}[htbp]
\begin{center}
\includegraphics[clip,width=9.5cm]{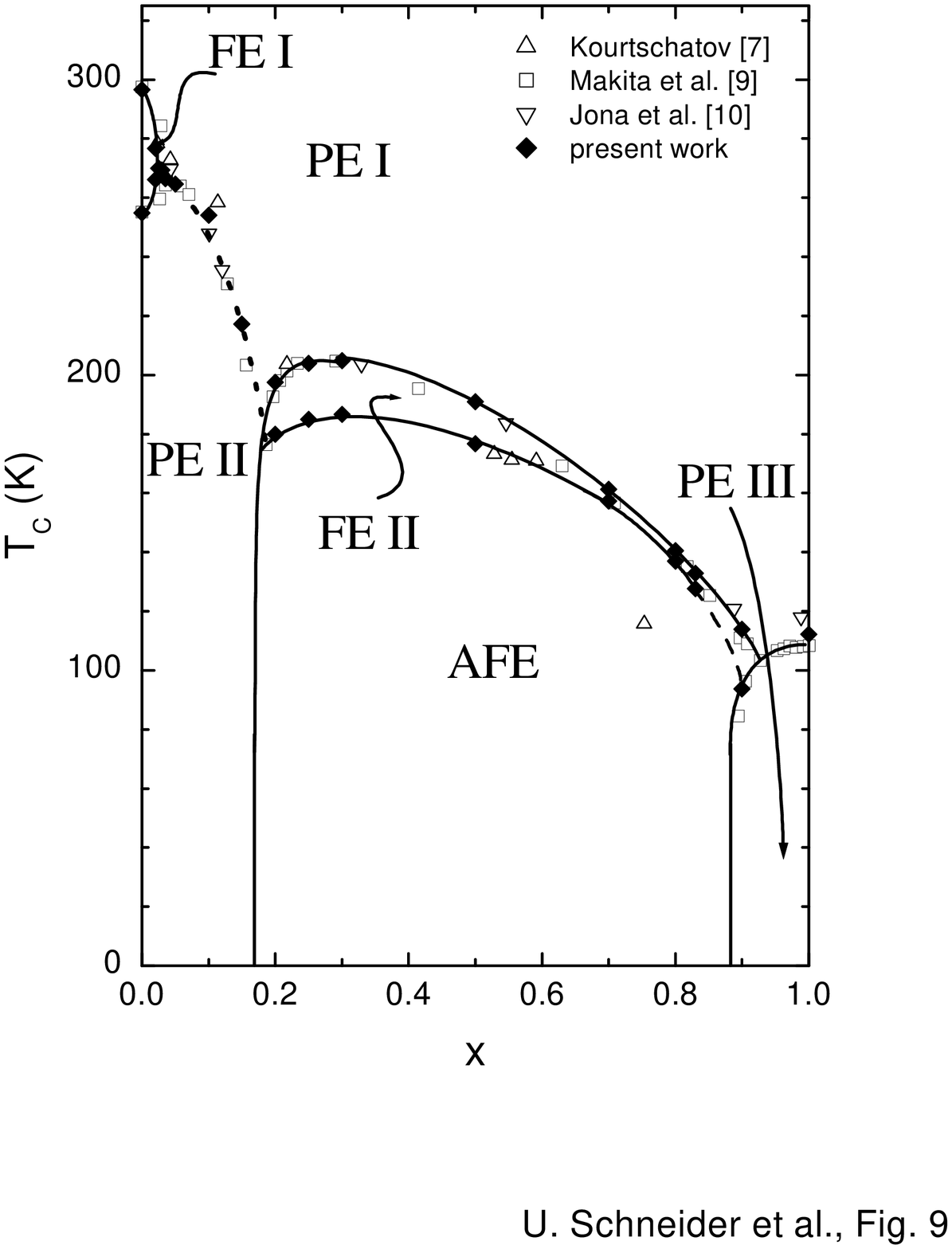}
\caption{Phase diagram of Na$($K$_{1-x}($NH$_4)_x)$C$_4$H$_4$O$_6\cdot 4$H$_2$O. The
filled symbols indicate transition temperatures determined in the present work except for
the range of $0.025<x<0.18$ where the dotted line follows the temperature of the flat,
broad maximum observed for these samples. The open symbols are taken from the
literature.} \label{fig:phases}
\end{center}
\end{figure}

Finally we will describe the modified phase diagram for the full ammonium concentration
range of $0<x<1$ in Figure~\ref{fig:phases}.
All previously published phase transitions are reproduced and refined on the basis of our
results. With decreasing temperature we observe a succession of
PE$\rightarrow$FE$\rightarrow$PE phases for $0<x<0.025$, denoted as PE I, FE I and PE II.
Above $x=0.025$ there is no indication for any ordering: the FE I phase vanishes
successively on increasing doping concentration; the dotted line in the phase diagram
indicates the position of a flat, broad maximum in $\varepsilon$'$(T)$. A possible
microscopic explanation for this feature is discussed in the following section. In the
range of $0.18<x<0.89$ however, our data justify a modification of the phase diagram:
while there is indeed a ferroelectric phase transition as reported before, this phase (FE
II) is followed by another r{\'e}gime with {\it anti}\,-ferroelectric correlations at
lower temperatures (AFE). We therefore introduce a new phase line in
Figure~\ref{fig:phases}. This result is supported by the fact that this behaviour was
reproduced in different samples and in measurements using two different experimental
setups. Above $x=0.89$ we observe a "polar" phase (PE III) where no order can be induced
electrically by fields below the threshold limit. The detailed physical nature of this
phase is still unresolved.

\section{MODEL CALCULATIONS}\label{sec:theory}

There has been a tremendous effort
\cite{blinc72,mitsui58,takagi72,blinc74,takagi75,sannikov79,kozlov88,okada89a,okada89b}
to explain the reason of the puzzling behaviour of the pure NaK-tartrate compound, namely
showing two second order transitions and a ferroelectric phase between them, taking into
account dipole-dipole and dipole-charge interaction between sublattices. From vibrational
spectroscopy of soft-mode behaviour \cite{kamba95,volkov99} there is strong evidence in
favour of the model of coupled oscillators of Reference~\cite{kozlov88,volkov99}. In the
present paper we apply the simple model of Blinc and \v{Z}ek\v{s} \cite{blinc72,blinc74}
which considers the interaction of the hydroxyl groups. The model, which is sometimes
called {\it pseudo-spin model}, is based on an asymmetric double-well potential with two
intertwined sublattices being mirror images of each other. Omitting a tunneling parameter
for the present case, because of the high temperatures at which the transition takes
place, the Hamiltonian of this arrangement may be expressed as follows:
\begin{align*}
 {\cal H}= &\frac{1}{2} \sum_{ij}\left[K_{ij}(S_{i1}^zS_{j1}^z+S_{i2}^zS_{j2}^z) + L_{ij}S_{i1}^zS_{j2}^z\right] \nonumber \\
           & - \Delta\sum_j(S_{j1}^z-S_{j2}^z)-2\mu\sum_j\left[E_{j1}(t)S_{j1}^z + E_{j2}(t)S_{j2}^z \right]
\end{align*}
where the $S_{ij}$ are {\it pseudo}\,spin-$\frac{1}{2}$ operators with the indices 1 and
2 denoting the two sublattices; the tensors $K_{ij}$ and $L_{ij}$ stand for the effective
interaction constants of the dipoles in the same or in the other sublattice. $\Delta$ is
a measure for the symmetry of the double wells and $\mu$ is the dipole moment interacting
with an external field. These assumptions and a mean-field approximation lead to a
self-consistent system of two equations. For a detailed derivation the reader may refer
to \cite{blinc74}. From this model we get the scalar parameters $K,L$ and $\Delta$ which
determine that the system is stable for a certain set of parameters
 \begin{center}
\begin{tabular}{rcl}
 for     &$ T<T_{c,1}$ or $T>T_{c,2}$ & with $\langle S_1^z\rangle = -\langle S_2^z\rangle$ \\[1ex]
 and for &$T_{c,1}\leq T\leq T_{c,2}$ & with $\langle S_1^z\rangle \neq -\langle S_2^z\rangle$
\end{tabular}
\end{center}
Here $L$ is the dominant term which leads to an antiparallel alignment of the sublattices
for all temperatures and all applied external fields. The asymmetry term $\Delta$ is also
responsible for an antiparallel tendency of the sublattices. The parameter $K$ describes
an antiparallel influence of overnext neighbours resulting in a ferroelectric tendency
for the whole system in the temperature range around the value of $K$.

Describing the physics of the model qualitatively one may explain the mechanism as
follows. The parts of the molecule now called "particles" which are responsible for the
polar order are located in double well potentials. There is no net polarization in the
ground state ({\it i.e.} the dipoles of both sublattices are opposite and equal); and
there are excited states with a finite dipole moment. Furthermore there may be an
interaction between the dipoles of the excited states which favours a parallel alignment.

At $T=0$ all molecules are in the ground state. Therefore there is no total polarization.
If the dipole-dipole interaction is large enough, the excitation energy of the particles
can be compensated by the gain in total energy due to the alignment of the dipoles. In
this scenario the state of ferroelectric order is favoured with respect to the total
energy of the system.

Now, the dipole-dipole interaction may be just below the critical value for this
scenario. For $T>0$ we have to take both the energy and the entropy into account: at a
finite temperature there is a probability for a certain number of particles to be in the
excited state ({\it i.e.} resulting in a net polarization). From this follows, that the
effective magnitude of the polarization is equal to the original magnitude multiplied by
this probability. On the other hand, the influence of the dipole-dipole interaction is
decreasing with rising temperature. Therefore, at $T>0$, a ferroelectric phase is only
possible if the probability for the excited states increases more rapidly than the
influence of the order decreases. At high temperatures the order will vanish in any case.

\begin{figure}[htbp]
\begin{center}
\includegraphics[clip,width=9.5cm]{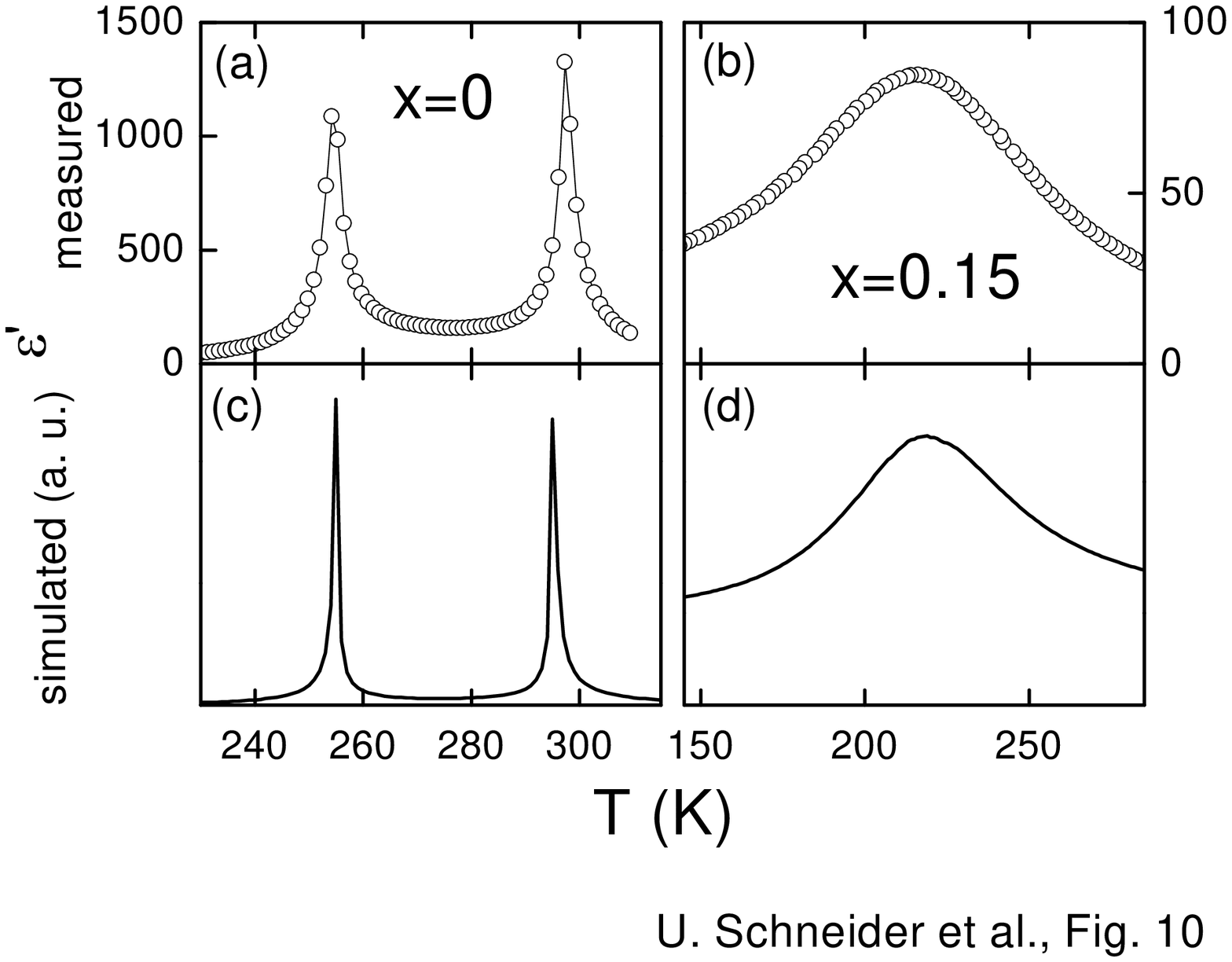}
\caption{Comparison of the measured data for the ammonium concentrations of $x=0$ (a) and
$x=0.15$ (b) with calculated curves according to the model of Blinc and \v{Z}ek\v{s}
\protect\cite{blinc72} in (c) and (d). The lines in (a) are drawn to guide the eye.}
\label{fig:theo}
\end{center}
\end{figure}

We will now apply the model to data of two ammonium concentrations: the sample of pure
NaK-tartrate ($x=0$) and the sample with $x=0.15$, the latter being in the range where no
FE ordering occurs any more. We take the data for the lowest frequency measured, namely
20\,Hz, to be close to the static limit of the theoretical model.
Figure~\ref{fig:theo}
shows that the theoretical result is in good qualitative agreement with our measurements.
We yield the parameters shown in Table~\ref{tab:1}.
\begin{table}[h]
\begin{center}
\begin{tabular}{|r|r|r|} \hline
 \,\,parameter\,\, & \,\,$x=0$\,\,        & \,\,$x=0.15$\,\,     \\ \hline
    \,\,K\,\,      & \,\,$271.275$\,K\,\, & \,\,$213.195$\,K\,\, \\ \hline
    \,\,L\,\,      & \,\,$5877.18$\,K\,\, & \,\,$4618.87$\,K\,\, \\ \hline
 \,\,$\Delta$\,\,  & \,\,$1701$\,K\,\,    & \,\,$1342$\,K\,\,    \\ \hline
\end{tabular}
\caption{Parameters for modeling the dielectric spectra for the two doping concentrations
$x=0$ and $x=0.15$ with the {\it pseudo-spin model} described in the text.} \label{tab:1}
\end{center}
\end{table}
From this table it is obvious that the set of parameters does not have to be changed very
much to describe the r{\'e}gime of suppressed FE order. For the description of the doped
sample the effective parameters $K$ and $L$, as well as the effective asymmetry parameter
$\Delta$ are lowered by about $27$\,\%.

\section{CONCLUSIONS}

We have presented results from low and high field frequency-dependent dielectric
measurements on 18 different ammonium doping concentrations of the compound
Na$($K$_{1-x}($NH$_4)_x)$C$_4$H$_4$O$_6\cdot 4$H$_2$O. Characteristic spectra of
concentrations representative for the various phases in the complex phase diagram were
shown. In the low field dielectric measurements a richness of relaxational features was
found over the whole doping range and systematically analyzed. The detected relaxation
processes were classified roughly into three groups. Presumably two of these groups
represent single particle intramolecular motions, whereas the third one is of collective
nature. From the hysteresis loops of the high field measurements we have provided
striking evidence for a new phase  showing anti-ferroelectric correlations in the doping
range of $0.18<x<0.89$. This finding is corroborated by the detection of an anomaly in
the temperature dependent dielectric permittivity obtained from the low field
measurements. A new phase boundary is introduced into the established phase diagram
showing the systematic doping dependence of the new phase. Finally we demonstrated for
the first time that the simple model by Blinc and \v{Z}ek\v{s} can be used to describe
not only the data on pure Rochelle salt, but the whole range up to an ammonium content of
$x=0.18$ at least qualitatively. This is a remarkable fact and may improve the
understanding of the re-entrant behaviour of the pure Rochelle salt vanishing upon
doping.

\section*{ACKNOWLEDGEMENTS}

We want to express our gratitude towards A. Maiazza at the Technische Universit\"{a}t
Darmstadt for growing the high-quality Rochelle salt crystals this work is based upon. We
are indebted to H. Ries for help with the high field non-linear measurements. One of the
authors (U.S.) also wants to acknowledge fruitful discussions with J. Petzelt and S.
Kamba at the Institute of Physics in Prague and A. Volkov at the General Physics
Institute in Moskow.

\end{document}